\documentclass[twocolumn,10pt]{revtex4-2}
\usepackage{graphicx}
\usepackage{amsmath}
\usepackage{txfonts}
\usepackage{textcomp}
\usepackage{color}

\begin{document}
\title{Plasmon-enhanced polarized single photon source directly coupled to an optical fiber}

\author{Masakazu Sugawara$^{1}$, Yining Xuan$^{1}$, Yasuyoshi Mitsumori$^{1, 2}$, Keiichi Edamatsu$^{1,*}$, and Mark Sadgrove$^{1, 3, \dagger}$}

\affiliation{$^1$RIEC, Tohoku University, Sendai, 980-8577, Japan\\
	$^2$Kitasato University, Kitasato University, Sagamihara, Kanagawa, 252-0373, Japan\\
    $^3$Tokyo University of Science, Shinjuku, Tokyo, 162-8601, Japan}

\email{$^*$eda@riec.tohoku.ac.jp}    
\email{$^\dagger$mark.sadgrove@rs.tus.ac.jp}

\begin{abstract}
A bright source of fiber-coupled, polarized single photons is an essential component of any realistic quantum network based on today's existing fiber infrastructure. Here, we develop a Purcell enhanced, polarized source of single photons at room temperature by coupling single colloidal quantum dots to the localized surface plasmon-polariton modes of single gold nanorods, combined on the surface of an optical nanofiber. A maximum enhancement of photon emission of 62 times was measured corresponding to a degree of polarization of 86 $\%$, and a brightness enhancement of four times in the fiber mode. Evanescent coupling of photons to the nanofiber guided modes ensures automatic coupling to a single mode fiber. Our technique opens the way to realizing bright sources of polarized single photons connected to fiber networks using a simple composite technique.
\end{abstract}
\maketitle

\section{Introduction}
The generation and storage of polarized single photons (``flying qubits") and their efficient coupling to existing fiber networks is an essential technology for the realization of quantum networks \cite{Kimblel2008, Lu, Uppu, Brekenfeld}. In addition, such photons should be generated at a high enough rate to keep pace with modern communication technologies in the gigahertz range. One research program which shows promise in achieving the above goals is the use of nano-waveguide based cavities to produce waveguide coupled single photons with a Purcell enhanced generation rate \cite{Okamoto2004, Iwase2010, Englund2015, Aharonovich2016, Schroder2016, Liu2018}. Nanowaveguide-based cavities are either automatically fiber coupled \cite{Nayak2014, Yalla2014, Schell2015, Li2017, Nayak2018} or can be coupled to optical fibers with high efficiency \cite{Tiecke2015}.

The requirements for such schemes may be split into three main parts: i) Efficient coupling of photons to the waveguide, ii) Enhancement of the rate of generated single photons and iii) Enhancement of photon degree-of-polarization (DOP). Photonic crystal-based nanowaveguide cavities are known to accomplish i) well, due to the overlap of the Purcell-enhanced cavity mode with the waveguide mode. However, with some notable exceptions \cite{Englund2015}, they achieve ii) to only a moderate degree, with enhancement factors of order 10 being common. More significantly, nano-waveguide-based cavities to date improve the degree of photon polarization only slightly, due to the mild cavity birefringence \cite{Yalla2014}. Nonetheless, photon polarization is essential for many photon based quantum communication schemes, suggesting that polarization enhancement is in vital need of attention.

The localized surface plasmon-polariton resonances (LSPRs) of metal nanostructures provide an important method of achieving polarization and brightness enhancement of single photon sources at room temperature. LSPRs function as Purcell regime cavities \textit{par excellence} with rapid decay rates and large emitter-resonator coupling strengths \cite{Tame2013, Bozhevolnyi2017, Bogdanov2019}. In addition, they exhibit very large birefringence, with orthogonally polarized modes having resonance wavelengths separated by hundreds of nanometers. The use of metal nano-structure LSPRs on nanowaveguides offers an overlapping set of advantages compared to waveguide-based cavities. Although i) coupling to the waveguide mode is not in general enhanced (due to the approximate point dipole nature of typical LSPRs), ii) brightness enhancement is relatively very large~\cite{schietinger2009plasmon, Yuan2009} and iii) room temperature polarization enhancement is unmatched~\cite{Unitt2005, kukushkin2014, Cadusch2015}. It is therefore clear that one path to room temperature, fiber-coupled polarized single photons requires a combination of waveguide-based cavities and LSPR based emitter enhancement. However, demonstrations of LSPR enhanced and polarized single photon emission on a nanowaveguide are lacking to date. 

Here, we developed an enhanced polarized single photon source at room temperature directly coupled to an optical fiber by depositing colloidal quantum dots (QDs) near single gold nanorods (GNRs) on the surface of an optical nanofiber. The spontaneous emission rate was  enhanced due to the LSPR of the GNRs with enhancement factors of up to 62. Moreover, due to the strong polarization anisotropy of the plasmonic resonance of the GNRs, we also measured an increase in the DOP to 86\% for photons emitted from the QD-GNR coupled system compared with a DOP of $\lesssim 50\%$ for QDs by themselves. Lastly, despite the fact that coupling efficiency to the fiber is not directly enhanced by this technique, a maximum enhancement of single photon intensity in the fiber modes of about 4 times was observed over the \emph{entire} QD bandwidth. This is in contrast to cavity based techniques, which typically only give strong enhancement for a small portion of the room temperature QD bandwidth. The above values were found to be in good agreement with simulation results.

\section{Physical principles and experiment}
\begin{figure*}[t]
	\centering
	\includegraphics[width=\linewidth]{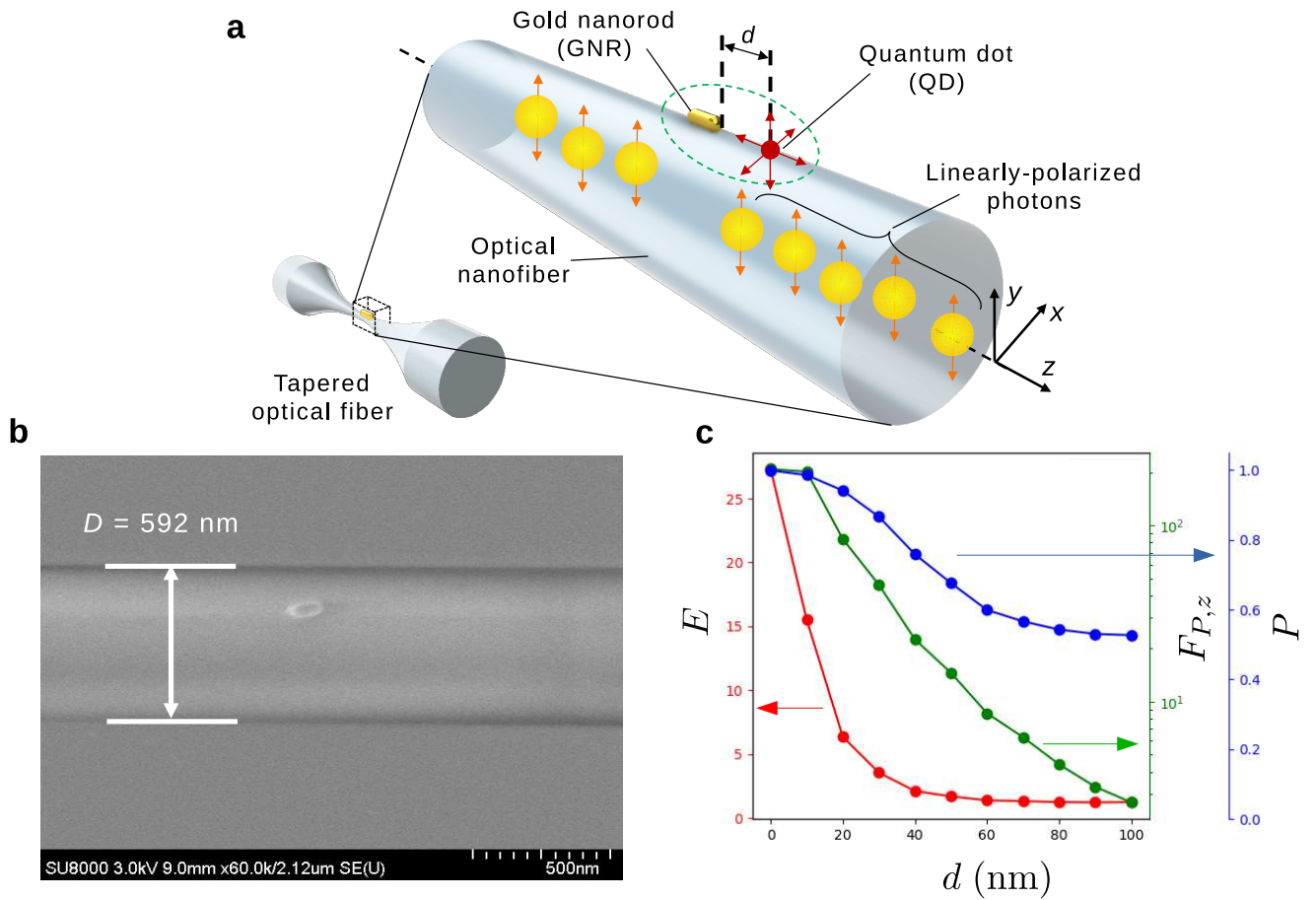}
	\caption{\textbf{Schematic illustration of the quantum-dot-gold-nanorod coupled system on an optical nanofiber}. \textbf{a}, Schematic illustration of the QD-GNR coupled system on an optical nanofiber. $d$ is the separation between the edge of the gold nanorod (GNR) and the colloidal quantum dot (QD) nanocrystal. \textbf{b}, Image of the GNR taken using a scanning electron microscope (SEM). \textbf{c}, Calculated intensity enhancement $E$, maximum Purcell enhancement factor $F_{P,z}$, and degree of polarization (DOP) $P$ for photons propagating in the guided mode of the optical nanofiber as a function of $d$ for a wavelength of 800 nm. }
	\label{fig:Concept}
\end{figure*}

\begin{figure}[t]
	\centering
	\includegraphics[width=\columnwidth]{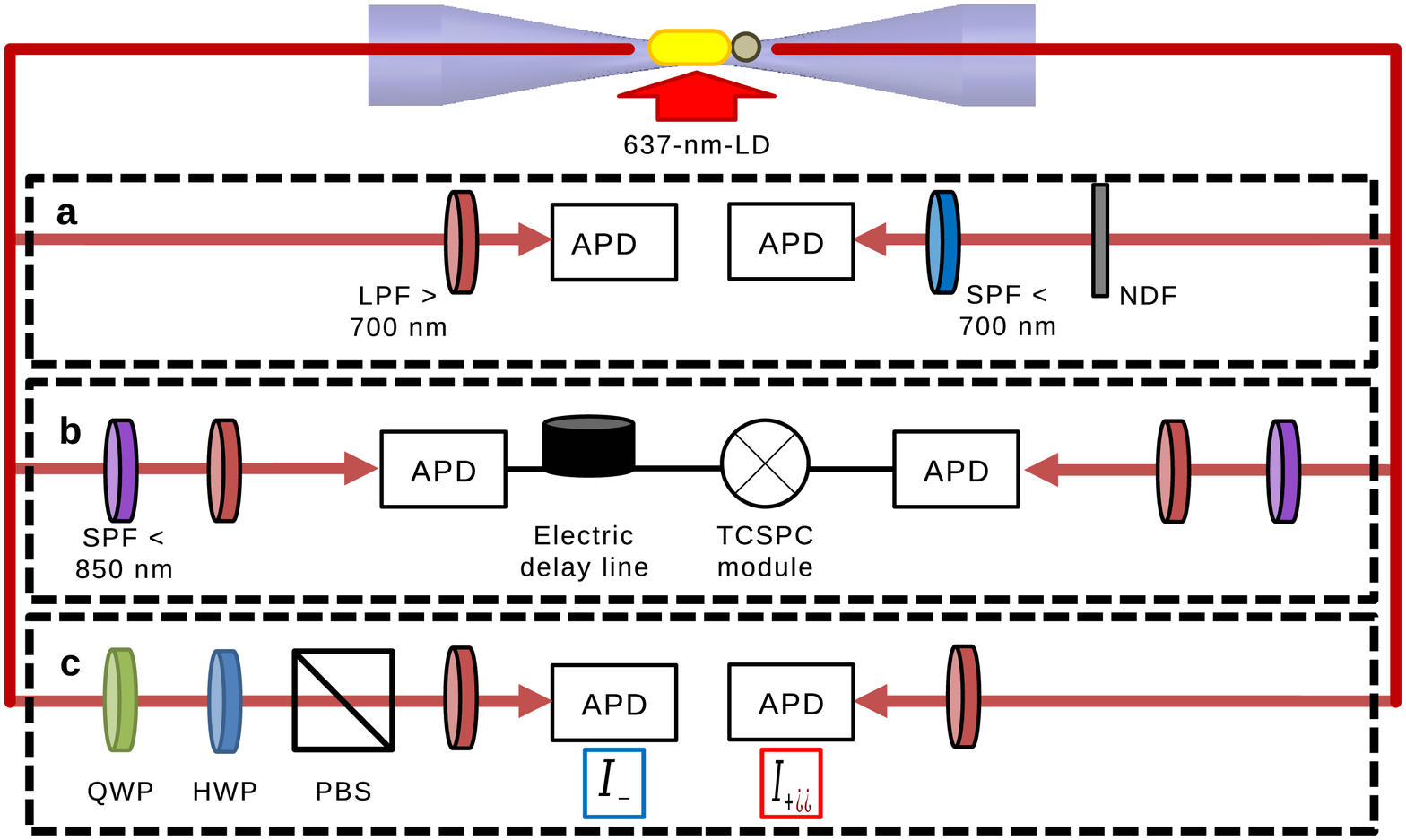}
	\caption{\textbf{Experimental setup}. The filter configuration for \textbf{a}, the fiber scanning measurement, \textbf{b}, the photon coincidence measurement, and \textbf{c}, the photon polarization measurement. Abbreviations denote the following: LD laser diode, LPF long-pass filter, APD avalanche photo diode, SPF short-pass filter, NDF neutral-density filter, TCSPC module time-correlated single photon counting module, QWP quarter wave-plate, HWP half wave-plate, PBS polarization beam splitter. We refer the signals detected at the APDs on the right (left) hand side to $I_{+}$ ($I_{-}$).}
	\label{fig:ExperimentalSetup}
\end{figure}

A schematic illustration of the experiment is shown in Fig. \ref{fig:Concept}\textbf{a}. The colloidal QD crystal emits single photons with a wavelength $\lambda_{\mathrm{QD}}$ and a spontaneous emission lifetime $\tau_{1}$ and a portion of the emitted photons are coupled into the guided mode of the optical nanofiber. If we place a GNR near the QD and the enhancement spectrum of the LSPR corresponding to the rod axis overlaps with the emission spectrum of the QD, the spontaneous emission of the QD will be greatly enhanced due to the Purcell effect. A scanning electon microscope (SEM) image of a single gold nanorod located on the surface of a nanofiber is shown in Fig. \ref{fig:Concept}\textbf{b}. We note that single QDs are difficult to image on the surface of an insulator such as the silica nanofiber using a SEM due to their small size and charge-up effects. 

Because the LSPR in the wavelength range of the QD has a single polarization,  spontaneous emission with a polarization lying along the GNR axis will be principally enhanced, leading to a large increase in the DOP of the emitted single photons~\cite{sonnichsen2002, Ming2009, Joos2018}. We calculated the expected values of the Purcell enhancement factor $F_{p,z}$ along with the DOP of the emitted photons $P$, and the intensity enhancement in the nanofiber guided modes $E$ using the finite-difference time-domain (FDTD) method. The results are shown in Fig. \ref{fig:Concept}\textbf{c}. In the simulations, the QD was modeled by a point dipole emitting over a range of wavelengths from 600 to 900 nm and the simulations were performed separately for three orthogonal dipole orientations. The nanofiber was modeled as a pure silica cylinder of diameter 530 nm, and the GNR was represented by a pure gold rod with hemispherical end caps, with a length of 160 nm and a diameter of 25 nm. For simplicity, the GNR was set to be parallel to the nanofiber axis, and positioned at the point $(x=0,y=530/2, z=0)$ nm. The results are plotted as a function of the distance $d$ between the edge of the GNR and the point dipole. The DOP was calculated using the expression
\[P = \dfrac{T_{y} + T_{z} - T_{x}}{T_{y} + T_{z} + T_{x}},\] 
where $T_{i}$ ($i = x, y, z$) is the proportion of the emitted light coupled to the nanofiber for a dipole polarized along the $i$ axis. For the case where the QD comes into contact with the GNR ($d$ = 0), $F_{p,z}$ is expected to exceed 200 and $P$ approaches unity. For completeness, we note that previous studies have found quenching behavior in the case where the emitter is in contact with metal nanostructures \cite{Anger2006}, and these effects are not included in the present simulations. The maximum Purcell factor $F_{p,z}$, fluorescence enhancement $E$ and the DOP $P$ are seen to fall to the levels seen for a dipole on a bare nanofiber as $d$ approaches 100 nm.

We now explain our sample preparation method. 
We first fabricated an optical nanofiber (average waist diameter 500 nm) from a commercially-available single-mode optical fiber (780HP, Thorlabs Inc.) using the heat and pull method \cite{Li}. The transmission through the fiber was over 90\% and the gradient of the fiber taper was sufficiently small that only the fundamental mode propagated for the wavelengths considered here. We used CdSeTe/ZnS core-shell type semiconductor colloidal QD crystals (Q21371MP, Thermo Fisher Scientific Inc.) as single photon emitters. These QDs have a nominal emission wavelength of 800 nm with a spectral bandwidth full-width at half-maximum (FWHM) of 50 nm. We note that the center wavelength can vary widely for the QDs used here and we observed QD spectra with center wavelengths between 760 nm and 800 nm for individual QDs. The centre wavelength of the LSPR of the GNRs (A12-50-800-CTAB-DIH-1-25, Nanopartz Inc.) was 800 nm with a spectral FWHM of 200 nm. 

We deposited the nanoparticles on the surface of the optical nanofiber as follows: First, we extracted several tens of \textmu L of the colloidal nanoparticle solution using a micro pipette. We then prepared a droplet of the particles at the tip of the pipette and brushed the droplet on the surface of the optical nanofiber several times. By introducing light from a laser diode (wavelength 637 nm) into the fiber, scattering from GNRs on the nanofiber surface could be seen confirming deposition. For the case of the deposition of the GNRs, we achieved single GNR depositions with high probabilities by de-aggregating GNRs in the solution with ultrasonic waves for 10 minutes just prior to deposition. 

After particle deposition, photoluminescence (PL) from the QDs and scattered light from GNRs deposited on the optical nanofiber was measured using the experimental setup as shown in Fig. \ref{fig:ExperimentalSetup}. The optical nanofiber was mounted on a three-axis computer-controllable translation stage. We excited the deposited particles with laser light (wavelength 637 nm) generated by a free-running semiconductor diode. The laser light was focused by an objective lens (Nikon, 20$\times$ magnification, NA = 0.3) and the fiber-lens distance was adjusted to achieve a FWHM spot size of 5 \textmu m. PL from QDs or scattered light from GNRs coupled to the fiber guided modes was detected using fiber coupled avalanche photo-diodes (APDs) after passing through the relevant filter system as illustrated in Fig. \ref{fig:ExperimentalSetup}. We label the signals detected at the right (left) hand side as $I_{+}$ ($I_{-}$) below. The light scattered or emitted by the deposited particles and coupled into the guided modes of the optical nanofiber passed through a free-space filtering system before being coupled to a multi-mode fiber and detected by APDs. We used three different filter configurations (\textbf{a}-\textbf{c} in Fig. \ref{fig:ExperimentalSetup}) depending on the type of measurement. 

 We evaluated the polarization response of the depositions on the nanofiber surface according to a method we developed previously \cite{Sadgrove2017,sugawara2020} to establish with high probability that the depositions were indeed single GNRs. 

 After confirmation of a single GNR deposition, we touched a droplet of the solution containing the QDs to the nanofiber at the position of the observed scattering points of the GNRs and held the droplet there for two minutes as 640 nm light was continuously  transmitted through the nanofiber mode. The solution of the QDs was diluted to 80 nmol/L using pure water. We note that this method of QD deposition was probabilistic and we were able to obtain QD-GNR coupled systems with a probability of about 13$\%$ per attempt. Assessment of the optimal parameters for achieving single QDs coupled to single GNRs with high probability using the above method is a topic of ongoing research.

To measure $P$, we used the filter configuration shown in Fig. \ref{fig:ExperimentalSetup}\textbf{c} and measured the polarization of photons coupled into the guided modes of the optical nanofiber. On the left hand side of the filter system, the photon polarization was first rotated as near as possible to the linear polarization basis using the QWP. We then measured the transmission through the HWP and PBS with respect to the angle of the HWP. On the right hand side of the filter system, only a 700-nm-LPF was inserted, allowing us to make a reference measurement of the emitted photon counts without polarization filtering. Calculating the time-averaged values of the fractions $I_{-}/I_{+}$ for each angle of the HWP, we could find the degree of photon polarization. In all measurements of $P$ shown in this paper, the error bars indicate one standard deviation of the fraction of transmission. We used a sinusoidal fit to the data points, and calculate the degree of polarization $P$ according to the following equation
\begin{equation}\label{eq:P}
P = \dfrac{\mathrm{max}(\langle I_{\mathrm{frac}}(\theta_{\mathrm{HWP}})\rangle) - \mathrm{min}(\langle I_{\mathrm{frac}}(\theta_{\mathrm{HWP}})\rangle)}{\mathrm{max}(\langle I_{\mathrm{frac}}(\theta_{\mathrm{HWP}})\rangle) + \mathrm{min}(\langle I_{\mathrm{frac}}(\theta_{\mathrm{HWP}})\rangle)},
\end{equation}
where $\langle I_{\mathrm{frac}}(\theta_{\mathrm{HWP}})\rangle = \langle I_{-}(\theta_{\mathrm{HWP}})/I_{+}(\theta_{\mathrm{HWP}})\rangle$ is the time-averaged transmission through the PBS with respect to the angle of HWP, $\theta_{\mathrm{HWP}}$. 

\section{Results}
\begin{figure*}[t]
	\centering
	\includegraphics[width=\linewidth]{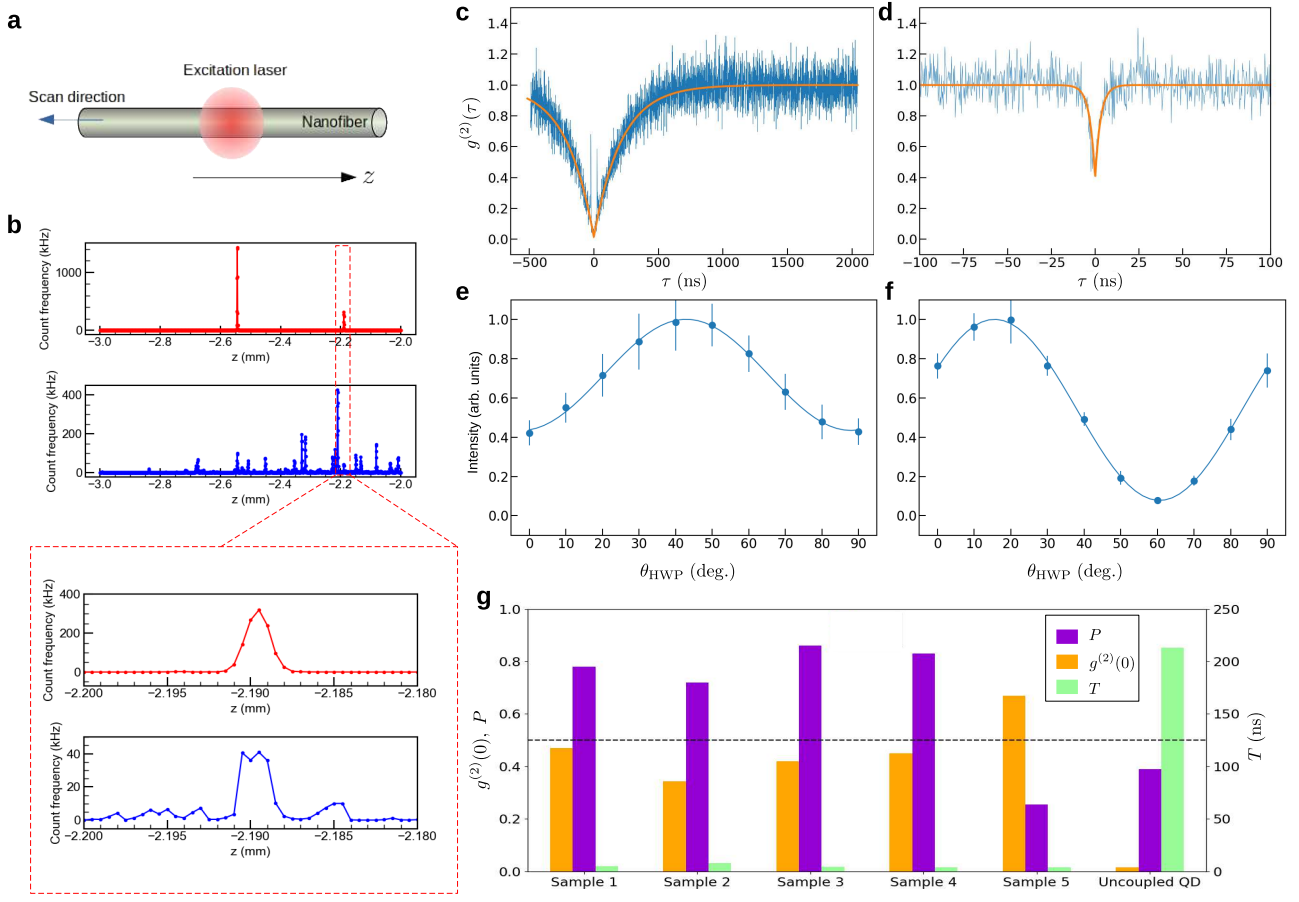}
	\caption{\textbf{Measurements for single gold rods and single QDs}. \textbf{a} Schematic depiction of the sweep 
	used to detect gold nanorods and QDs. \textbf{b} Results of a sweep showing detection of gold nano-rods (red points, upper panel) and QDs (blue points, lower panel). The inset shows a zoomed-in image revealing that gold rod and QD peaks are overlapped near $z=-2.19$. \textbf{c,d} Intensity correlation $g^{(2)}(\tau)$ for the case of QDs uncoupled and coupled to a GNR respectively. \textbf{e,f} Measurements of the degree of polarization $P$ for the cases shown in \textbf{c} and \textbf{d} respectively.  \textbf{g} Summary showing $g^{(2)}(0)$ (orange bars), degree of polarization $P$ (purple bars) and rise time $T$ (green bars) for five different samples where coupling between a QD and a GNR was present.  Sample 3 corresponds to the data shown in \textbf{d} and \textbf{f}. The final position shows the case for a single quantum dot without any coupling to a GNR and corresponds to the data shown in \textbf{c} and \textbf{e}.}
	\label{fig:Expt1}
\end{figure*}

We first measured the distribution of the deposited particles using the filter configuration \textbf{a} as shown in Fig.~\ref{fig:ExperimentalSetup}. By sweeping the excitation laser light along the surface of the optical nanofiber, as depicted in Fig.~\ref{fig:Expt1}\textbf{a}, and detecting the light from the deposited particles coupled into the guided modes of the optical nanofiber, we could determine the positions of either type of particle deposited at each position along the optical nanofiber surface. The filter system on the left hand side collected only the PL signal from deposited QDs by filtering out the excitation light coupled into the nanofiber guided modes using a long-pass filter with a cut-on wavelength of 700 nm. As for the right hand side filter system, because the scattered light from the GNRs was too strong for the APDs, the scattered light was attenuated by inserting neutral-density filters (NDFs). The measured particle distribution is shown in Fig. \ref{fig:Expt1}\textbf{b}. The PL (scattered) light signal is displayed in the top (bottom) panel. Zooming in on the bright peak from the GNR around z = -2.19 mm (inset, Fig. \ref{fig:Expt1}\textbf{b}), the two peaks from the GNR and the QD deposition are seen to  clearly overlap within the $\sim$ 5$\;\mu$m spot size of the excitation laser and thus the possibility of coupling between a GNR and a QD exists in this case.

We now discuss measurements made for depositions where a QD and a GNR were found at the same position. First, we consider the second-order correlation function $g^{(2)}(\tau)$. Measurement of $g^{(2)}(\tau)$ provides two principle pieces of information about the coupled QD-GNR light source. First, a value $g^{(2)}(0)<0.5$ indicates single photon dominant light emission. Second, the rise time $T$ of $g^{(2)}(\tau)$ as $|\tau|$ increases gives the emitter lifetime. In particular, normalized photon coincidences from a single quantum emitter show antibunching given by
\[g^{(2)}(\tau) = 1-\mathrm{exp}(-\dfrac{|\tau|}{T})\]
where~\cite{Yalla2012,Akimov2007}, 
\begin{equation}
T = \left( \alpha P_{\mathrm{exc}} + 1/\tau_{1}\right)^{-1},
\label{Eq:T}
\end{equation}
 $\alpha P_{\mathrm{exc}}$ is the power dependent excitation rate and $\tau_1$ is the intrinsic decay time of the emitter. Coupling of a QD to a GNR gives rise to a decrease in $T$ due to the Purcell effect, and thus the value 
 of $T$ relative to the uncoupled value $T_0$ gives information about the degree of fluorescence enhancement at a given 
 excitation power. 
 
Experimental measurements were made by correlating the PL signals taken from both ends of the optical fiber. Note that the optical nanofiber itself plays the role of a non-polarizing beam splitter in this case. The filter configuration on both sides was identical. Photon coincidences as a function of interval $\tau$ were recorded using a time correlated single photon counting (TCSPC) module (TimeHarp200, PicoQuant). 
850-nm short-pass filters (SPFs) were inserted to reduce the effect of weak photon emission from the APDs themselves~\cite{Tao:2005Photon}. 
We first focus on measurements of $g^{(2)}(\tau)$ at a given excitation power. In particular, measurements of antibunching were made for excitation powers between 2 and 3 $\mu$W for the case of an uncoupled QD and for five different coupled QD-GNR samples. In this narrow power range, there is little power induced variation of $T$, 
and differences reflect the strength of coupling between the QD and the GNR. 

Figure~\ref{fig:Expt1}\textbf{c} shows the case for an uncoupled quantum dot on an optical nanofiber with an estimated diameter of 530 nm.   Here, $g^{(2)}(0)$ was found to be 0.02 and $T$ was found to be 210 ns. These values are in good agreement with typical values found in previous studies of single QDs on optical nanofibers~\cite{Yalla2012}.
Figure~\ref{fig:Expt1}\textbf{d} shows a measurement of $g^{(2)}(\tau)$ for the QD-GNR sample which gave the largest overall enhancement in our experiments. The antibunching dip is seen to have a value $g^{(2)}(0)=0.42$. Comparison of the results shown in Figs.~\ref{fig:Expt1}\textbf{c} and \textbf{d} clearly reveals large Purcell enhancement, with  $T$ reduced to 4.4 ns in Fig.~\ref{fig:Expt1}\textbf{d}.

Next, we consider measurements of the degree of polarization $P$ of the fluorescence. Figure~\ref{fig:Expt1}\textbf{e} shows 
a measurement of visibility made using the setup shown in Fig.~\ref{fig:ExperimentalSetup}\textbf{c} for the same uncoupled QD on a nanofiber measured in Fig.~\ref{fig:Expt1}\textbf{c}. The corresponding degree of polarization was found to be $P=0.39$.  Figure~\ref{fig:Expt1}\textbf{f} shows similar measurements made for the same sample measured in Fig.~\ref{fig:Expt1}\textbf{d}. We see a large enhancement of the degree of polarization in this case up to $P=0.86$.
These observations are consistent with Purcell enhancement by the highly anisotropic plasmon resonance of the gold nanorod.

Finally, we summarize the results for for five samples (Sample 1 through Sample 5) in which a QD was found to be deposited close to a GNR on a nanofiber, and compare the results with an uncoupled QD as shown in the bar chart of Fig.~\ref{fig:Expt1}\textbf{g}. The results indicate coupling between a QD and a GNR according to the greatly reduced values of $T$ compared to the uncoupled QD case. Note that Sample 3 corresponds to the measurements shown in Figs.~\ref{fig:Expt1}\textbf{d} and \textbf{f}, and was the sample for which the smallest value of $T$ and the largest value of $P$ was measured. The lowest antibunching dip was seen for Sample 2 with $g^{(2)}(0)=0.34$, clearly in the regime of single photon dominant emission.

It is interesting to note that the only sample, Sample 5, for which $g^{(2)}(0)$ was clearly larger than 0.5 (indicating two or more QDs were contributing strongly to the fluorescence) also showed a decreased value of $P$, in comparison to the other samples, where $P$ was found to be enhanced relative to the uncoupled case. The relatively small value of $P$ in this case suggests that only one of the two QDs in Sample 5 was significantly coupled to the GNR.

In order to evaluate the Purcell factor, it is necessary to evaluate the ratio of the QD-GNR system decay rate $\gamma$ to that of a QD in free space $\gamma_0$, i.e. $F_{P,z} = \gamma / \gamma_0 = \tau_0 / \tau_1$. Note that we take $\tau_0 / \tau_1$ to correspond to \emph{the maximum Purcell factor} over the broad enhancement region where the QD spectrum overlaps with the plasmonic resonance of the GNR. This is because the \emph{fastest} decay channel experienced by the QD defines the sharpness of the rise time of $g^{(2)}(\tau)$ measured over the emission bandwidth. Furthermore, for simplicity, we will take $\tau_0$ to be the value measured for the uncoupled QD. Strictly speaking, the nanofiber itself imparts a Purcell enhancement to the QD decay, however, the polarization average of the enhancement is close to unity.

\begin{figure*}[t]
	\centering
	\includegraphics[width=\linewidth]{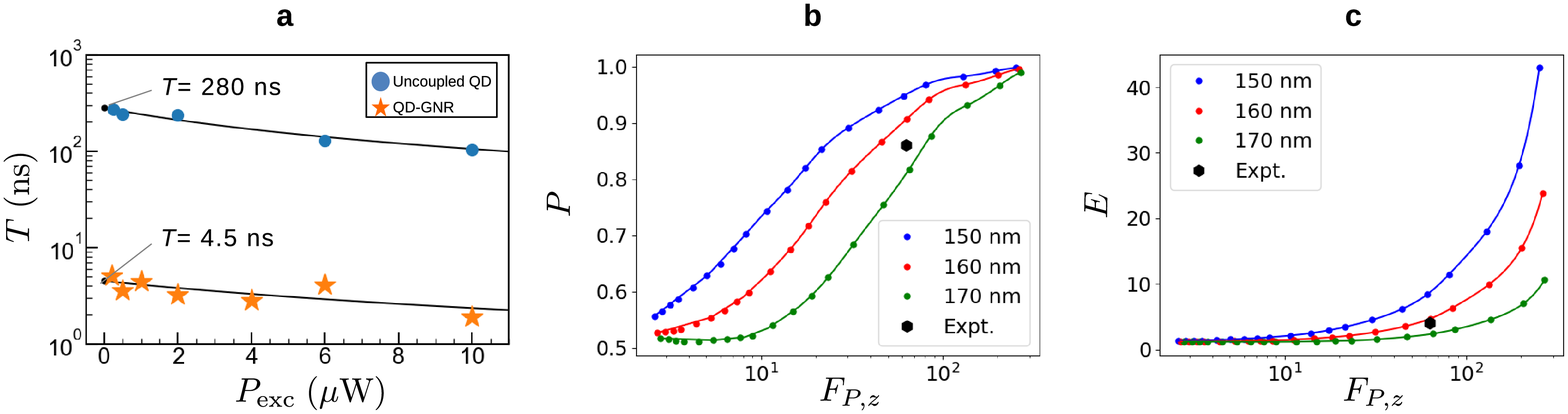}
	\caption{\textbf{Evaluation of the peak Purcell factor}. \textbf{a} Measurements of rise time $T$ as a function of excitation power for the case of a single QD uncoupled to a GNR (blue circles) and a QD coupled to a GNR (yellow stars). Solid lines show a fit to the data using Eq.~\ref{Eq:T}.  \textbf{b} Simulation results showing the polarization $P$ vs. the peak Purcell factor $F_{P,z}$ for the nano-rod lengths indicated in the legend. The experimentally measured value is shown by a black hexagon.  \textbf{c} Simulation results showing the intensity enhancement $E$ vs. the peak Purcell factor $F_{P,z}$ for the nano-rod lengths indicated in the legend. The experimentally measured value is shown by a black hexagon.}
	\label{fig:Expt2}
\end{figure*}

To measure the maximum Purcell factor produced by our technique, we measured $T$ at various excitation powers for both the uncoupled QD measured above, and Sample 3 where the best coupling was observed.
Figure~\ref{fig:Expt2}\textbf{a} shows the power dependence of $T$ in the cases of a QD uncoupled to a GNR (blue circles) and for Sample 3 (yellow stars). By fitting the experimental data by Eq.~\ref{Eq:T}, we evaluated the spontaneous emission lifetimes of the single QD and the QD-GNR coupled system as $\tau_0=\tau^{QD}_{1} =$ 280 ns and $\tau^{QD-GNR}_{1} =$ 4.5 ns, respectively. Therefore, the Purcell enhancement factor can be estimated to be $\tau_0/\tau^{QD-GNR}_{1} = 62$. Note that this estimate of the Purcell factor relies on the measured value of $\tau_0$ being representative of the entire QD ensemble. In the current experiment, we checked $\tau_0$ for only one uncoupled QD. However, thorough studies of the same QDs coupled to optical nanofibers have been performed in previous work~\cite{Yalla2012}, allowing us to assign an uncertainty to $\tau^{QD}_{1}$ of 30$\%$. Using this value, a rough estimate of the uncertainty of the Purcell enhancement factor measured here was found to be $F_{P,z}=62 \pm 26$.

We also made measurements of the maximum photon count rate achieved in the case of the uncoupled QD and Sample 3. 
These rates were found to be 40 kHz and 150 kHz in the case of uncoupled QD and GNR-QD systems respectively.
Thus, the enhancement of fluorescence coupled to the fiber modes was found to be $E=3.8$. The reason that this value is so much smaller than the Purcell enhancement itself is that the enhancement reflects the \emph{average} increase in coupled intensity over the QD emission bandwidth, rather than its maximum as in the case of the Purcell factor measurements. Furthermore, by making the $z$-directed component of the dipole emission dominant, the Purcell effect actual \emph{reduces} the coupling efficiency of fluorescence to the nanofiber since the $z$-dipole couples the most weakly out of the three principle dipole alignments $x$, $y$ and $z$. More specifically, we numerically calculated the coupling efficiency from the linearly-polarized point dipole oriented at 23 degrees (as in the experiment for Sample 3) into the guided modes of the optical nanofiber using FDTD simulations and the calculated coupling efficiency was 6.8 \% whereas the calculated averaged coupling efficiency for the case of a randomly-polarized point dipole was 17.7 \%. Correcting for this drop in the coupling efficiency suggests that the real enhancement of 
fluorescence intensity is more than 10 times.

In any case, we note that an overall enhancement of fluorescence intensity of $\approx4$ is actually impressive relative to most conventional cavity based schemes for Purcell enhancement. This is because at room temperature, the narrow spectral width of a typical cavity mode overlaps with only of order 10$\%$ of the emission bandwidth of a QD~\cite{Yalla2014,Nayak2018}, leading to little real fluorescence enhancement when averaged over wavelength. Indeed, most quoted enhancement factors in the literature merely give the enhancement achieved at the cavity resonance wavelength. In this sense, while the absolute number is small, the value of $E$ achieved here also represents a practical improvement over many previous experiments. Of course, adding a cavity to the present QD-GNR system will allow large increases in the photon collection efficiency.

\section{Discussion and conclusion}
Measurement of the maximum Purcell factor as shown in Fig.~\ref{fig:Expt2}\textbf{a} allows us to make a detailed comparison of the results with predictions based on the FDTD method. We calculated the degree of polarization $P$, 
intensity enhancement $E$ and maximum Purcell factor $F_{P,z}$ for a number of separations $d$ between the QD and the GNR.
Note that both $P$ and $E$ are averages over the probability distribution defined by the normalized QD spectrum (center wavelength 760 nm, FWHM 50 nm to match the case of sample 3). Although the GNRs used in our experiment had a nominal length of 160 nm, in reality a spread in values of about 10\% is present in the sample, and the scanning electron microscope measurements are at the same level of accuracy. For this reason, we also performed simulations for the case where the GNR length was 150 nm and 170 nm for comparison. For simplicity we oriented the GNR along the $z$-axis. Simulation results for $P$ vs. $F_{P,z}$ and $E$ vs $F_{P,z}$ are shown in Figs.~\ref{fig:Expt2}\textbf{b} and \textbf{c} respectively, with symbols as indicated in the legend.
The experimental values for Sample 3 are shown as a black hexagon in both cases. We see remarkably good agreement between
the measured values and the simulation predicted values. In particular, the experimental values in the case of both $P$ and $E$ lie close to the curve for a rod length of $L=160$ nm, i.e. the nominal GNR length, and, more precisely, lie between the curves for $L=160$ nm and $L=170$ nm. The value of $d$ corresponding to these results is $\sim$ 25 nm in both cases. The agreement seen between simulations and experimental measurements suggests that the coupled QD-GNR system behaves close to ideally, i.e. the effects of quenching, etc, are not dominant possibly due to the distance between the QD and the GNR.

On the other hand, the occurence of relatively large values of $g^{(2)}(0)\lesssim 0.5$ for the coupled QD-GNR systems measured here is a curious feature only partly explained by the limitations of our measuring equipment when $T$ becomes small. Further experimental and theoretical studies are necessary to establish the detailed, single photon emission dynamics 
of QD-GNR coupled systems~\cite{Waks}.

In conclusion, we have developed a polarized rate-enhanced single photon source coupled to an optical fiber at room temperature by using coupling between QD nanocrystals and the LSPRs of anisotropic gold nanoparticles. If a more deterministic deposition method can be developed to control the position between the QDs and the nanoparticles, even larger photon emission enhancement and photon polarization enhancement can be expected. Aside from applications to quantum communication, we also note that the system considered here provides a promising platform for quantum plasmonics research, due to the convenience of the nanofiber substrate which also doubles as a collection device for emitted photons.

\section*{Acknowledgments}
This research was partially supported by MEXT Quantum Leap Flagship Program (MEXT Q-LEAP) Grant Number
JPMXS0118067581 and JSPS KAKENHI Grant Number JP20H01831. Mark Sadgrove acknowledges support from the Nano-Quantum Information Research Division of Tokyo University of Science. This research was partially performed using the facilities of the Fundamental Technology Center, Research Institute of Electrical Communication, Tohoku University.
\section*{Author contributions}
Masakazu Sugawara manufactured the samples, performed the experiments, developed the theory, analyzed the data, and wrote the first draft of the paper. X.Y. contributed to perform the experiments. Y.M. contributed to the experimental procedure and co-supervised the experiment. K.E. contributed to the experimental procedure, data interpretation and co-supervised the experiment. Mark Sadgrove contributed to the theory, experimental procedure, data analysis and co-supervised the overall project. All authors discussed the paper and contributed to revisions.

\end{document}